\documentclass[preprint,amsmath,amssymb,superscriptaddress,showpacs]{revtex4-1}
 
\usepackage[utf8]{inputenc}
\usepackage{amsmath}
\usepackage{latexsym}
\usepackage{amsfonts}
\usepackage{epsfig}
\usepackage{psfrag}
\usepackage{graphicx}
\usepackage{hyperref}
\usepackage{color}

\definecolor{black}{rgb}{0,0,0}
\definecolor{blue}{rgb}{0,0,1}
\definecolor{green}{rgb}{0,1,0}
\definecolor{red}{rgb}{1,0,0}
\definecolor{brown}{rgb}{0.4,0.2,0}
\definecolor{darkgreen}{rgb}{0,0.7,0}

\renewcommand{\vec}[1]{\boldsymbol #1}

\newcommand{\bea}{\begin{eqnarray}}
\newcommand{\ea}{\end{eqnarray}}
\newcommand{\eea}{\end{eqnarray}}

\usepackage[percent]{overpic}
 \newlength{\imagewidth}

\begin{document}
\setlength{\imagewidth}{0.5 \linewidth}

\title{Highly excited electronic image states of metallic nanorings}
\author{Christian Fey}
\email[]{christian.fey@physnet.uni-hamburg.de }
\affiliation{Zentrum f\"ur optische Quantentechnologien, Luruper Chaussee 149, 22761 Hamburg, Universit\"at Hamburg, Germany}
\author{Henrik Jabusch}
\affiliation{Zentrum f\"ur optische Quantentechnologien, Luruper Chaussee 149, 22761 Hamburg, Universit\"at Hamburg, Germany}
\author{Johannes Knörzer}
\altaffiliation[New address: ]{Max-Planck-Institut für Quantenoptik, Hans-Kopfermann-Str. 1, 85748 Garching, Germany}
\affiliation{Zentrum f\"ur optische Quantentechnologien, Luruper Chaussee 149, 22761 Hamburg, Universit\"at Hamburg, Germany}
\author{Peter Schmelcher}
\email[]{peter.schmelcher@physnet.uni-hamburg.de }
\affiliation{Zentrum f\"ur optische Quantentechnologien, Luruper Chaussee 149, 22761 Hamburg, Universit\"at Hamburg, Germany}
\affiliation{The Hamburg Centre for Ultrafast Imaging, Luruper Chaussee 149, 22761 Hamburg, Universit\"at Hamburg, Germany} 

%\date{\today}

\begin{abstract}
We study electronic image states around a metallic nanoring and show that the interplay between the attractive polarization force and a repulsive centrifugal force gives rise to Rydberg-like image states trapped several nanometers away from the surface.
The nanoring is modeled as a perfectly conducting isolated torus whose electrostatic image potential is derived analytically. The image states are computed via a two-dimensional finite-difference scheme as solutions of the effective Schrödinger equation describing the outer electron subject to this image potential.
These findings demonstrate not only the existence of detached image states around nanorings but allow us also to provide general criteria on the ring geometry, i.e. the aspect ratio of the torus, that need to be fulfilled in order to support such states.
\end{abstract}

\maketitle

%%%%%%%%%%%%%%%%%%%%%%%%%%%%%%%%%%%%%%%%%%%%%%%%%%%%%%%%%%%%%%%%%%%%%%%%%%%%%
\section{Introduction}
Image potential states are single-electron excitations trapped outside the crystal surface of a metal by the attractive interaction between the electron and the polarizable bulk \cite{echenique_existence_1978,echenique_theory_1989,echenique_decay_2004}. At large distances $z$ from the surface the interaction is governed by the classical electrostatic image potential which scales in the limit of an infinitely extended flat surface as $\propto 1/z$. In contrast, at short distances (a few Bohr radii) the specific microscopic material properties and many-body interactions become important \cite{bardeen_image_1940,lang_theory_1973,jung_self-consistent_2007}. 

On a structural as well as on a formal level, this situation is reminiscent of the physics of the outer valence electron in an alkali atom. Accordingly, the spectrum of image states above flat surfaces can be described very accurately by a Rydberg series with adapted quantum defects \cite{echenique_theory_1989}. Experimentally observed image states localize typically 10 to 100 Bohr radii above the surface with lifetimes that may extend to several ps \cite{echenique_decay_2004}. Studying the dynamics of these states via time-resolved two-photon photoemission spectroscopy has led to an improved understanding of scattering processes occurring at surfaces \cite{bovensiepen_electron_2010}.

In contrast to image states above flat surfaces, tubular image states (TIS) occur around curved surfaces like metallic nanotubes and are characterized by an angular momentum number $l$ \cite{granger_highly_2002,segui_tubular_2012}. 
For high angular momenta (typically $l\geq 7$) the interplay between the repulsive centrifugal barrier and the long-range part of the image potential forms stable radial potential wells that
support detached image states localized at distances of 10-50 nm from the surface. This trapping mechanism leads to enhanced lifetimes of these states which have been estimated to 1 ns-1 $\mu$s \cite{segal_ultraslow_2005}.
Since their prediction in 2002 \cite{granger_highly_2002} TIS have attracted much interest which led to numerous subsequent studies. This includes on the one hand explorations of novel setups like TIS at finite segmented nanowires \cite{segal_shaping_2004,knorzer_control_2015}, parallel nanowires \cite{segal_electric_2004}, nanowire lattices \cite{segal_tunable_2005} and fullerene buckyballs \cite{gumbs_image_2006}. On the other hand, this comprises also refined studies of the image force at the short-range scale for particular microscopically modeled semiconducting or conducting surfaces \cite{zamkov_image-potential_2004, gumbs_comparison_2005, segui_tubular_2012}. While low-$l$ TIS have been observed experimentally via two-photon photoemission at free standing multi-walled carbon nanotubes \cite{zamkov_time-resolved_2004} and via scanning tunneling microscopy/spectroscopy at supported double-walled carbon nanotubes \cite{schouteden_probing_2010}, an experimental detection of high-$l$ TIS has not yet been reported.

In the present paper we analyze high-$l$ TIS around nanorings. These TIS resemble those in Ref. \cite{granger_highly_2002} but are trapped at curved nanowires having both ends connected. To this aim we construct an exact expression for the electrostatic image potential and analyze subsequently electronic eigenstates bound by the competition between the image force and a repulsive centrifugal force. Although the angular momentum around the wire is not conserved we show that characteristic properties of these states can be derived from an effective radial potential. 

The paper is organized as follows. In Sec. \ref{sec:image_potential} we present the underlying electrostatic problem. In Sec. \ref{sec:electrostatic_problem} the image potential is set up in toroidal coordinates. In Sec. \ref{sec:image_pot_analysis} we analyze an exemplary nanoring and point out general scaling properties of the image potential. Sec. \ref{sec:potential_states} deals with image states trapped in the potential. In Sec. \ref{sec:finite_diff} we describe our numerical approach which is employed in Sec. \ref{sec:image_states_results} to compute image states. Finally, the states are analyzed and interpreted by means of an effective radial potential in Sec. \ref{sec:effective_radial_potential}. Our conclusions are provided in Sec. \ref{sec:conclusions}. 

\section{Nanoring Image Potential}
\label{sec:image_potential}
\subsection{The electrostatic problem}
\label{sec:electrostatic_problem}

\begin{figure}[h]
\includegraphics[width=0.35 \textwidth]{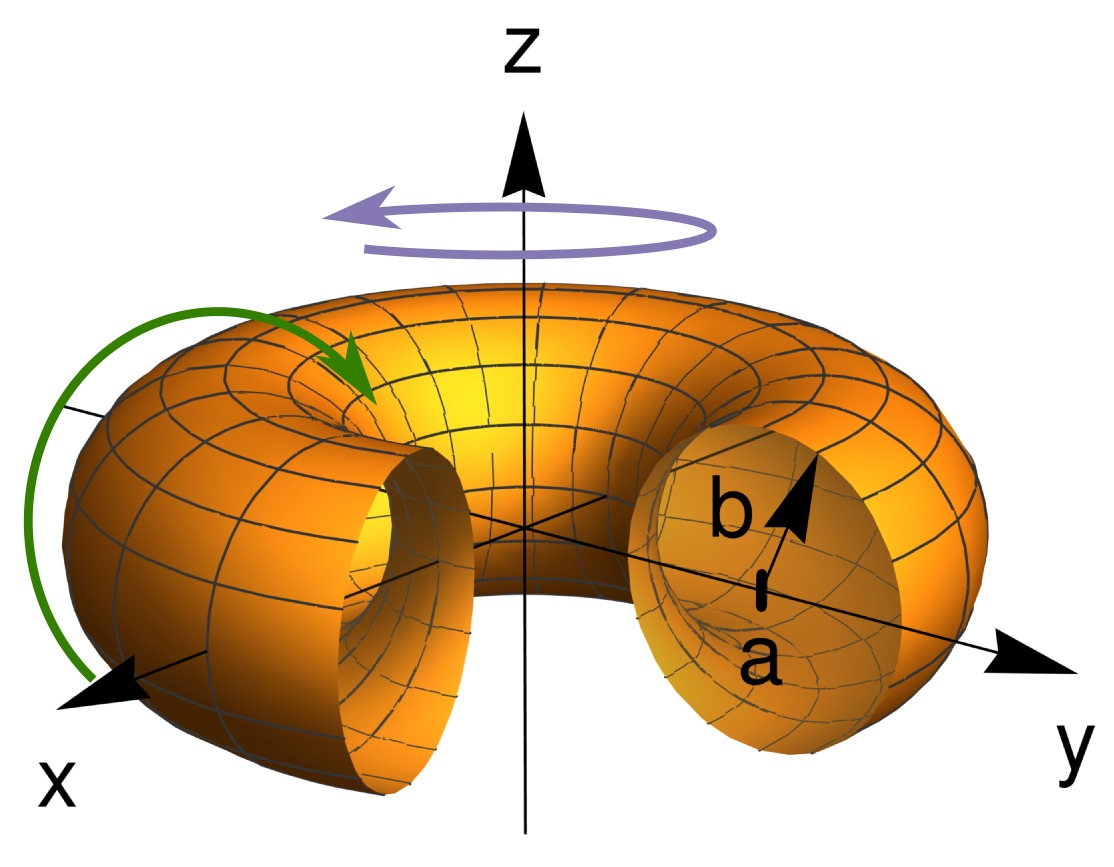}
\caption{Sketch of a torus with major radius $a$ and minor radius $b$. The colored arrows indicate the so-called toroidal (violet) and poloidal (green) direction. }
\label{fig:torus}
\end{figure}

We consider an electron at position $\vec{r}_e$ in the proximity of a conducting nanoring whose surface is parametrized as a torus with major radius $a$ and minor radius $b$ for $a>b$, see Fig. \ref{fig:torus}. 
If the position of the electron is fixed, the total electrostatic potential at any point $\vec{r}$ outside the ring can be expressed as $\Phi_\text{tot}(\vec{r};\vec{r}_e)= \Phi_\text{coul}(\vec{r};\vec{r}_e)+\Phi_\text{ind}(\vec{r};\vec{r}_e) $ where $\Phi_\text{coul}(\vec{r};\vec{r}_e)=-1/|\vec{r}-\vec{r}_e|$ is the Coulomb potential and $\Phi_\text{ind}(\vec{r};\vec{r}_e)$ is the induced potential which results from the polarization of the nanoring (for convenience we work here in atomic units). 
The boundary conditions for an ideal conductor require that the total electrostatic potential is constant for all points $\vec{r}_s$ lying on the surface of the conductor, i.e. $\Phi_\text{tot}(\vec{r}_s;\vec{r}_e)=V_0$. For neutral, i.e. non-charged, conductors this constant $V_0$ is a priori not known and needs to be determined by employing the additional condition that the induced potential $\Phi_\text{ind}(\vec{r};\vec{r}_e)$ is source-free, in contrast to grounded nanostructures ($V_0=0$) that have in general non-source-free $\Phi_\text{ind}(\vec{r};\vec{r}_e)$. Here we focus exclusively on neutral nanorings. In this case one can show that the energy needed to separate the electron and the nanostructure is given by $V_\text{im}(\vec{r}_e)= 1/2 \cdot \Phi_\text{ind}(\vec{r}_e;\vec{r}_e)$, which is called the image potential \cite{taddei_subtleties_2009}.

As for metallic cylinders \cite{granger_highly_2002}, the image potential for tori can be expressed analytically, too. To this aim we introduce toroidal coordinates $\xi$, $\eta$, $\varphi$ which are linked to cylindrical coordinates $\rho$, $z$, $\varphi$ via
\begin{equation}
\rho=\frac{c \sinh \xi}{\cosh \xi-\cos \eta}
\end{equation}
 and
\begin{equation}
z=\frac{c \sin \eta}{\cosh \xi- \cos \eta}
\end{equation}
with $c=\sqrt{a^2-b^2}$. In the new toroidal coordinates the torus surface is defined by $\xi=\xi_0$ with $\cosh\xi_0=a/b$.
Employing these relations, the image potential of the neutral torus can be constructed by combining classical electrostatic results obtained for conducting tori in different external field configurations \cite{scharstein_electrostatic_2005,smythe_static_1989,smythe_solutions_1974} and reads
\begin{equation}
V_\text{im}(\xi,\eta)=\frac{1}{2\pi c}(\cosh\xi-\sinh\eta) \left[\frac{s_1^2(\xi,\eta)}{s_1(0,0)}-s_2(\xi)\right] 
\label{eqn:vim_torus_neutral}
\end{equation}
where 
\begin{equation}
s_1(\xi,\eta)= \sum_{n=0}^\infty \epsilon_n \frac{Q_{n-1/2}(\cosh \xi_0)}{P_{n-1/2}(\cosh \xi_0)} P_{n-1/2}(\cosh \xi) \cos (n \eta)
\label{eqn:sum1}
\end{equation}
and
\begin{equation}
s_2(\xi)= \sum_{n=0}^\infty \sum_{m=0}^\infty \epsilon_n \epsilon_m (-1)^m \frac{\Gamma(n-m+1/2)}{\Gamma(n+m+1/2)} \frac{Q^m_{n-1/2}(\cosh \xi_0)}{P^m_{n-1/2}(\cosh \xi_0)} \left[P^m_{n-1/2}(\cosh \xi)\right]^2 .
\label{eqn:sum2}
\end{equation}
Here $P_{n-1/2}$, $Q_{n-1/2}$, $P^m_{n-1/2}$ and $Q^m_{n-1/2}$ are associated Legendre functions and $\epsilon_n=2-\delta_{0n}$ with the Kronecker delta $\delta_{0n}$. The detailed derivation of $V_\text{im}(\xi,\eta)$ is given in Appendix \ref{sec:imagepotential_derivation}.  
While the sum $s_1(\xi,\eta)$ converges for all $\xi \geq \xi_0$, the second sum $s_2(\xi)$ diverges for $\xi \to \xi_0$. This agrees well with the fact that the image potential of any ideal conductor will become arbitrarily large for a decreasing distance to its surface. To evaluate (\ref{eqn:vim_torus_neutral}) numerically we check the convergence of the sums $s_1(\xi,\eta)$ and $s_2(\xi,\eta)$ at each coordinate $\{\xi,\eta\}$ by increasing the range of summation until a desired precision is achieved.     

\subsection{Analysis of the image potential}
\label{sec:image_pot_analysis}
\begin{figure}[h]
\includegraphics[width=0.49 \textwidth]{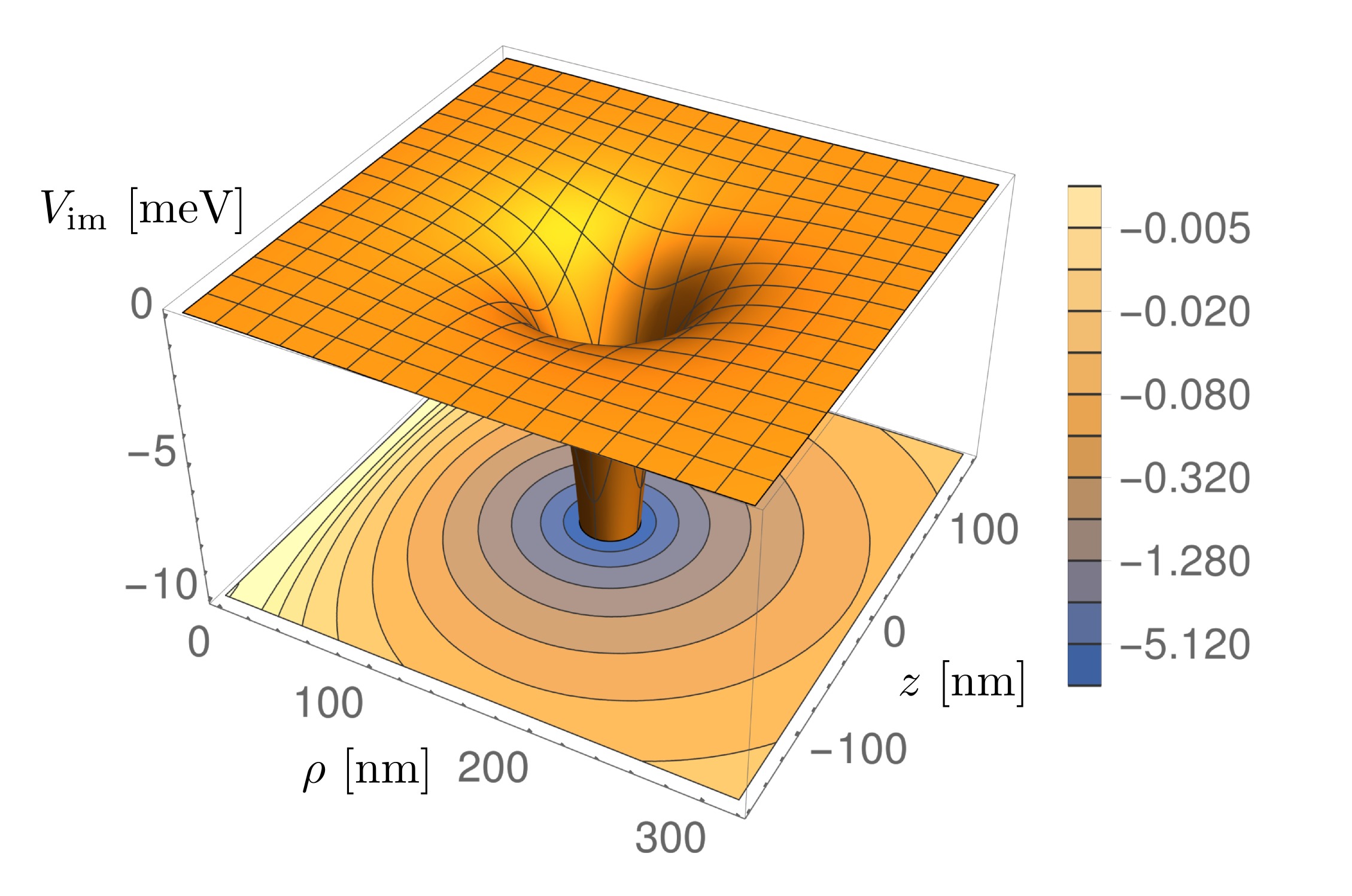}
\caption{Image potential $V_\text{im}(\rho,z)$ for an electron at a metallic nanoring with $a=3000$ ($\sim 159 \text{ nm}$) and $b= 30$ ($\sim 1.59 \text{ nm}$). A logarithmically scaled contour plot of $V_\text{im}(\rho,z)$ is shown (lower part of the figure) to illustrate the anisotropy of the potential.}
\label{fig:vim_torus_3d}
\end{figure}

As an illustrative example we consider the image potential of a ring with $a=3000$ ($\sim 159 \text{ nm}$) and $b= 30 $ ($\sim 1.59 \text{ nm}$). This length $b$ as well as the 
circumference $2 \pi a$ are comparable to the dimensions of single and multi-walled nanotubes with radii ranging from 0.68 nm to 7.1 nm and lengths ranging from 200 nm to 1600 nm which have been studied in the context of TIS \cite{granger_highly_2002,zamkov_image-potential_2004,segal_shaping_2004,knorzer_control_2015}. 
In Fig. \ref{fig:vim_torus_3d} we depict the resulting image potential in cylindrical coordinates. The plot illustrates the attraction of the electron towards the torus surface which is a circle of radius $b$ in the $\rho$-$z$ plane. Close to the surface the potential becomes nearly isotropic with respect to the tube center $\{\rho=a,z=0\}$ due to its divergent behavior. At larger distances from the surface, however, the anisotropy is clearly visible.   

\begin{figure}[h]
\includegraphics[width=0.49 \textwidth]{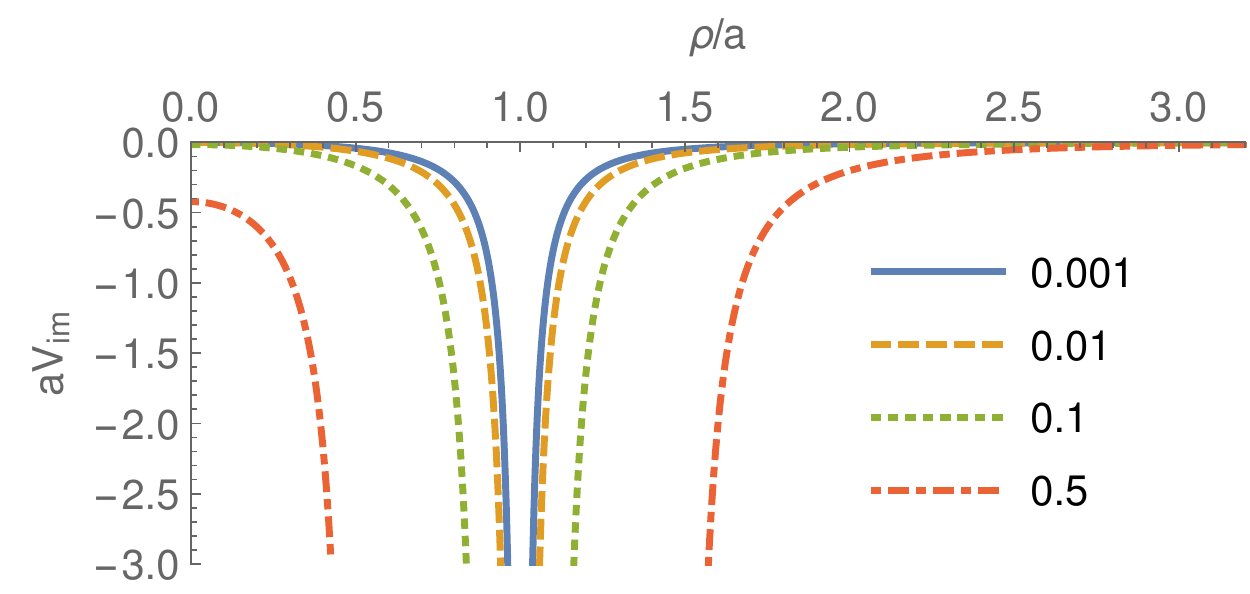}
\caption{Radial cuts of the rescaled image potential $a V_\text{im}(\rho,z=0)$ as a function of the rescaled radial distance $\rho/a$ for different ratios of the torus radii $b/a$ as stated in the legend. }
\label{fig:vrescale}
\end{figure}

General properties of nanorings with different parameters $a$ and $b$ can be illustrated by defining
the rescaled quantities $\tilde{\rho}=\rho/a$, $\tilde{z}=z/a$ and $\tilde{V}_\text{im}(\tilde \rho, \tilde z)=a V_\text{im}(\rho,z)$. The resulting rescaled potential $\tilde{V}_\text{im}(\tilde{\rho},\tilde{z})$ depends then only on the aspect ratio $b/a$. Radial cuts of this rescaled
potential are presented in Fig. \ref{fig:vrescale} for several $b/a$.    
As can be seen the ratio $b/a$, firstly, determines the positions of the torus surface $\tilde{\rho}=1\pm b/a$ where the potential curves diverge, and, secondly, it controls the asymmetry around $\tilde{\rho}=1$, which becomes clearly visible for the largest $b/a=0.5$.

What is not obvious from Fig. \ref{fig:vrescale} is that also the scaling properties of the image potential depend significantly on $b/a$. As we will discuss in the next section, the scaling is crucial for the capability of nanorings to support detached TIS. To analyze this property we compare the nanoring potential to a power-law with exponent $\alpha$ which is typical for many image potentials, e.g. at flat surfaces ($\alpha=-1$) or at grounded metallic spheres ($\alpha=-2$). 

\begin{figure}[h]
\includegraphics[width=0.49 \textwidth]{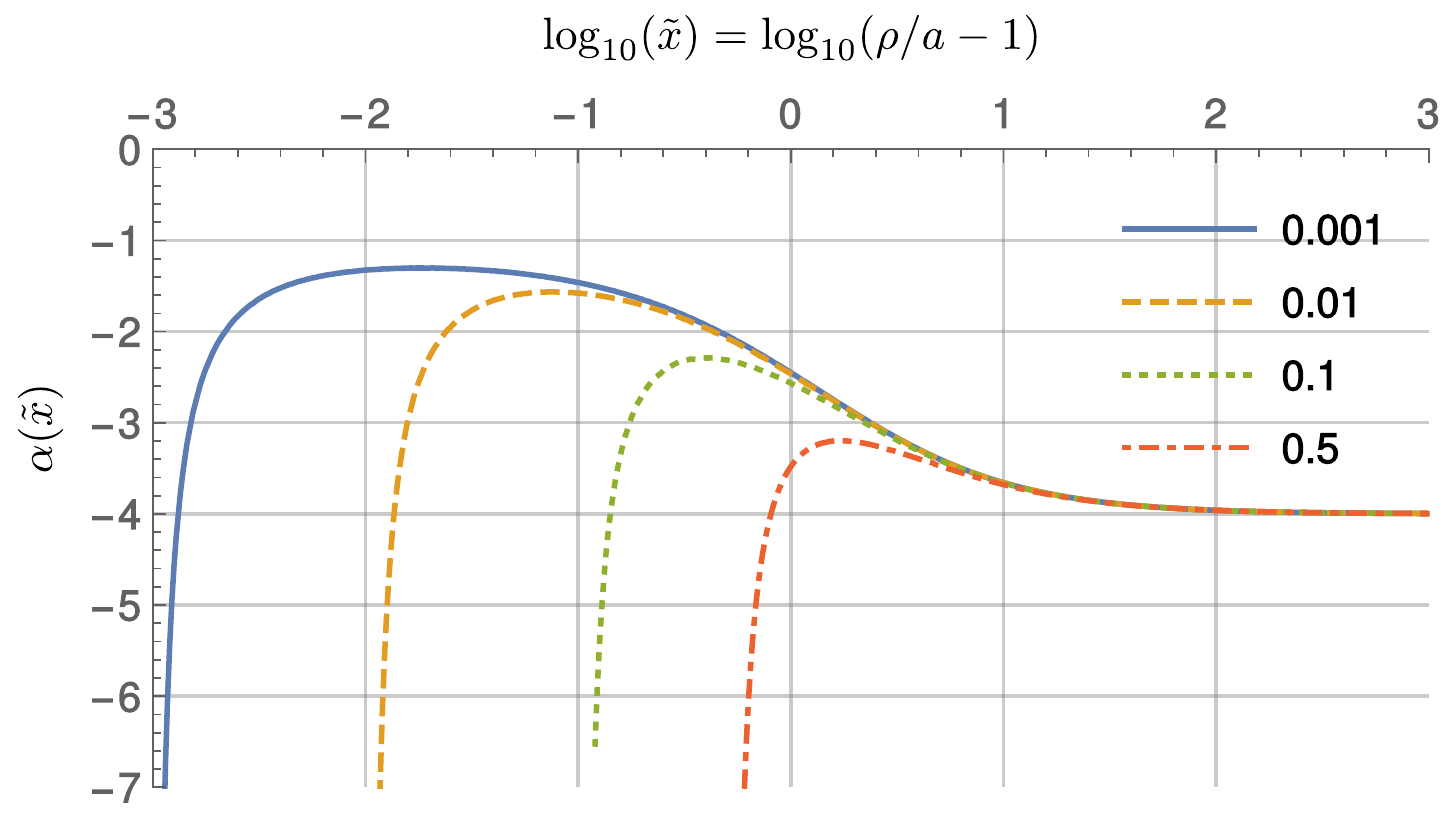}
\caption{Local exponent $\alpha(\tilde{x})$ as defined in (\ref{eqn:alpha}) in dependence of the rescaled radial distance $\tilde{x}$ to the tube center ($\rho=a$,$z=0$) for different ratios of the torus radii $b/a$.}
\label{fig:exponent}
\end{figure}

For simplicity we restrict this analysis here only to a radial cut of the potential in the region $\rho>a$.
By introducing the rescaled radial distance of the electron to the tube center $\tilde{x}= \tilde \rho-1$ the potential can be rewritten as $\tilde{V}^\text{rad}_\text{im}(\tilde{x})=\tilde{V}_\text{im}(\tilde{\rho},0)$. 
The local exponent of the potential is then defined by
\begin{equation}
\alpha(\tilde{x})= \frac{d}{d \ln \tilde{x}}  \ln \left (-\tilde{V}^\text{rad}_\text{im}(\tilde{x})\right)
\label{eqn:alpha}
\end{equation}
which is the slope in a double logarithmic plot of  $\tilde{V}^\text{rad}_\text{im}(\tilde{x})$. This exponent is depicted in Fig. \ref{fig:exponent} for different ratios $b/a$. All curves converge for large $\tilde{x}$ to $\alpha(\tilde{x})=-4$, which is the expected long-range scaling for the interaction of a point charge and a polarizable neutral finite object (monopole-induced-dipole-interaction). At short distances $\alpha(\tilde{x})$ diverges as expected for $\tilde{x} \to b/a$. In between these limits there is an intermediate regime where the exponent approaches a maximal value. The smaller the ratio $b/a$ the larger is this maximal value, e.g. one has $\alpha(\tilde{x})< -2$ if $b/a \geq 0.1$. In other words, only sufficiently thin nanorings have spatial regions where the radial image potential decreases slower than $\tilde{x}^{-2}$.

\section{Toroidal Image Potential States}
\label{sec:potential_states}
\subsection{Computational approach}
\label{sec:finite_diff}
Although electronic excitations at metal surfaces are in general a complex many-body phenomenon, image potential states have been described in many situations very successfully within a simple one-electron picture \cite{echenique_existence_1978,echenique_decay_2004, granger_highly_2002}. In this model one solves the Schrödinger equation of the single outer electron subject to the image potential. As the image potential expresses only the electrostatic energy of the system it is clear that dynamical corrections are required when retardation effects or collective excitations of the metal (e.g. plasma oscillations) become important  \cite{mills_image_1977,gumbs_comparison_2005}.         

The nanoring image potential $V_\text{im}(\rho,z)$ is rotationally symmetric around the $z$-axis and we solve the Schrödinger equation in cylindrical coordinates.  
The stationary solution with energy $E_{m,n}$ can be written as an eigenstate of the angular momentum operator $\hat{L}_z$ 
\begin{equation}
\Psi_{m,n}(\rho,z,\varphi)=\frac{u_{m,n}(\rho,z)}{\sqrt { 2 \pi \rho}}  e^{im\varphi} 
\end{equation}
with angular momentum $m$. The separation of the angular degree of freedom leads then to the two-dimensional stationary Schrödinger equation
\begin{equation}
\left[-\frac{1}{2}\left(\frac{\partial^2}{\partial \rho^2}+ \frac{\partial^2}{\partial z^2}\right)+ V^\text{eff}_\text{m}(\rho,z)-E_{m,n}\right]u_{m,n}(\rho,z)=0
\label{eqn:schrod_eqn_2d}
\end{equation}
for the wave function
$u_{m,n}(\rho,z)$ with an effective potential
\begin{equation}
V^\text{eff}_\text{m}(\rho,z)= \frac{m^2-1/4}{2\rho^2}+ V_\text{im}(\rho,z) 
\label{eqn:v_eff}
\end{equation}
that includes a centrifugal barrier depending on $m$. The second quantum number $n$ thus labels the different eigenstates and energies for fixed $m$.

Eq. (\ref{eqn:schrod_eqn_2d}) formally resembles a two-dimensional Schrödinger equation in Cartesian coordinates with the additional restriction that $\rho$ needs to be positive.
We compute the energetically lowest eigenstates by representing $u_{m,n}(\rho,z)$ as a vector on an equidistant rectangular grid with $N$ gridpoints $\{\rho_i,z_i\}$ for $ 1 \leq i\leq N$. In this representation $V_m^\text{eff}$ becomes a diagonal matrix and the derivative operators $\partial ^2/ \partial \rho^2$ and $\partial ^2/ \partial z^2$ can be expressed as sparse matrices by employing the finite-difference approximation (for our calculations we use a 7-point-stencil for each dimension).

To compute detached TIS with this method, two modifications of the original physical system are necessary. Firstly, we impose hard-wall boundary conditions at the borders of the grid which have no effect on bound states if the grid is chosen large enough (see below). Secondly, we introduce a cut-off energy $V_\text{min}$ to regularize the divergence at the surface and replace $V_\text{im}(\rho,z)$ by $V_\text{min}$ if $V_\text{im}(\rho,z)<V_\text{min}$. Among all computed states $u_{m,n}$ we then select states which are insensitive to the cut-off as well as to the boundary conditions. This method works well for wave functions localized in regions with $\rho \gg 1$ but it should be mentioned for completeness that hard-wall boundary conditions may in general be inappropriate for states having $m=0$ which are allowed to have finite values $\Psi_{0,n}(\rho,z)$ for $\rho \to 0$ and which are therefore not part of our analysis.  

\subsection{Numerical results}
\label{sec:image_states_results}
\begin{figure}[h]
\includegraphics[width=0.98 \textwidth]{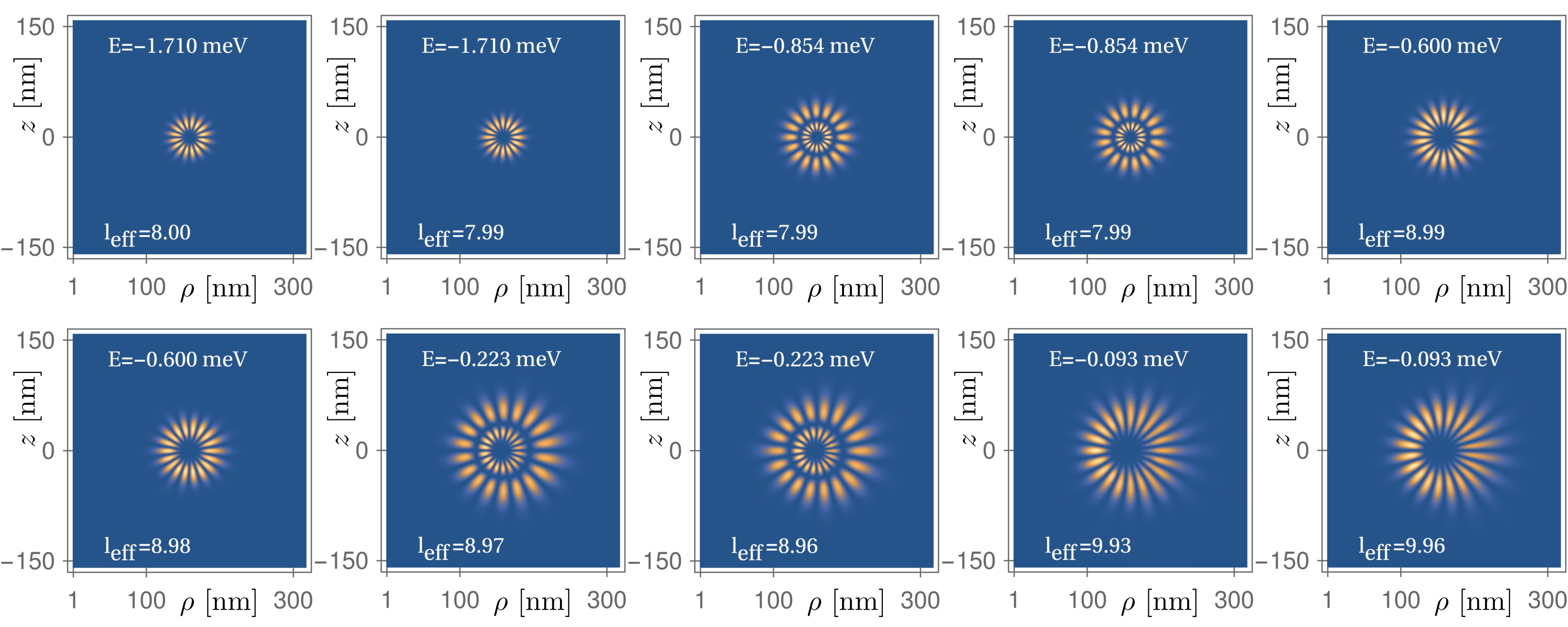}
\caption{Probability density $|u_{1,n}(\rho,z)|^2$ of the first ten TIS having the lowest probability  in the inner region of the nanoring with $a=3000$ ($\sim 159 \text{ nm}$) and $b= 30$ ($\sim 1.59 \text{ nm}$). Additionally, their energy $E$ and poloidal angular momentum $l_\text{eff}$ are displayed.}
\label{fig:denplots_torus}
\end{figure}

To illustratively exemplify the existence of TIS at nanorings we employ our computational approach to determine image potential states at the nanoring with $a=3000$ and $b= 30$. We choose $m=1$, the grid is built up by $1100 \times 1100$ grid points and we set $V_\text{min}\approx 4 \text{ eV}$ such that the image potential is cut off only in regions closer than 2 Bohr radii to the surface. In Fig. \ref{fig:denplots_torus} we show the ten states having the lowest integrated probability to be found in the inner torus region ($(\rho-a)^2+z^2\leq b^2$). This probability is on the order of $10^{-6}$.   

These states have a pronounced circular symmetry around the tube center $\{\rho=a,z=0\}$ and resemble in the transversal $\rho$-$z$-plane strongly high angular momentum states around nanotubes \cite{granger_highly_2002,segui_tubular_2012}. The main differences are, however, that the circular symmetry is not perfect (as can be clearly seen from the higher excited states in Fig. \ref{fig:denplots_torus}) and that the TIS are periodically extended along the toroidal $\varphi$-direction (not visible in Fig. \ref{fig:denplots_torus}).

To quantify the circular motion we define the angular momentum operator $\hat{l}=-i [ (\rho-a) \partial_z -z \partial_\rho ]$ which is the generator of rotations along the so-called poloidal direction, see Fig. \ref{fig:torus}.
For each state Fig. \ref{fig:denplots_torus} contains the expectation value of the squared angular momentum in terms of an effective angular momentum quantum number $l^2_\text{eff}=\left<\hat{l^2}\right>$. The numerical accuracy of $l_\text{eff}$ depends on our spatial resolution (here $\Delta \rho=\Delta z \approx 7 $) and can for this example be estimated to $\Delta l_\text{eff}\approx 0.02$.   

For the presented states the number $l_\text{eff}$ lies always close to an integer value counting the number of maxima of the electronic wave function in the $\rho$-$z$-plane, i.e. half of the maxima in the density plots.
States having the same number of radial and angular nodes appear as energetically nearly degenerate pairs (differences below 1$\,\mu$eV) and have a different parity along the $z$-axis. In the limit of an exact circular symmetry the degeneracy would become exact and $l_\text{eff}$ would become an integer. 

\begin{figure}[h]
\includegraphics[width=0.9 \textwidth]{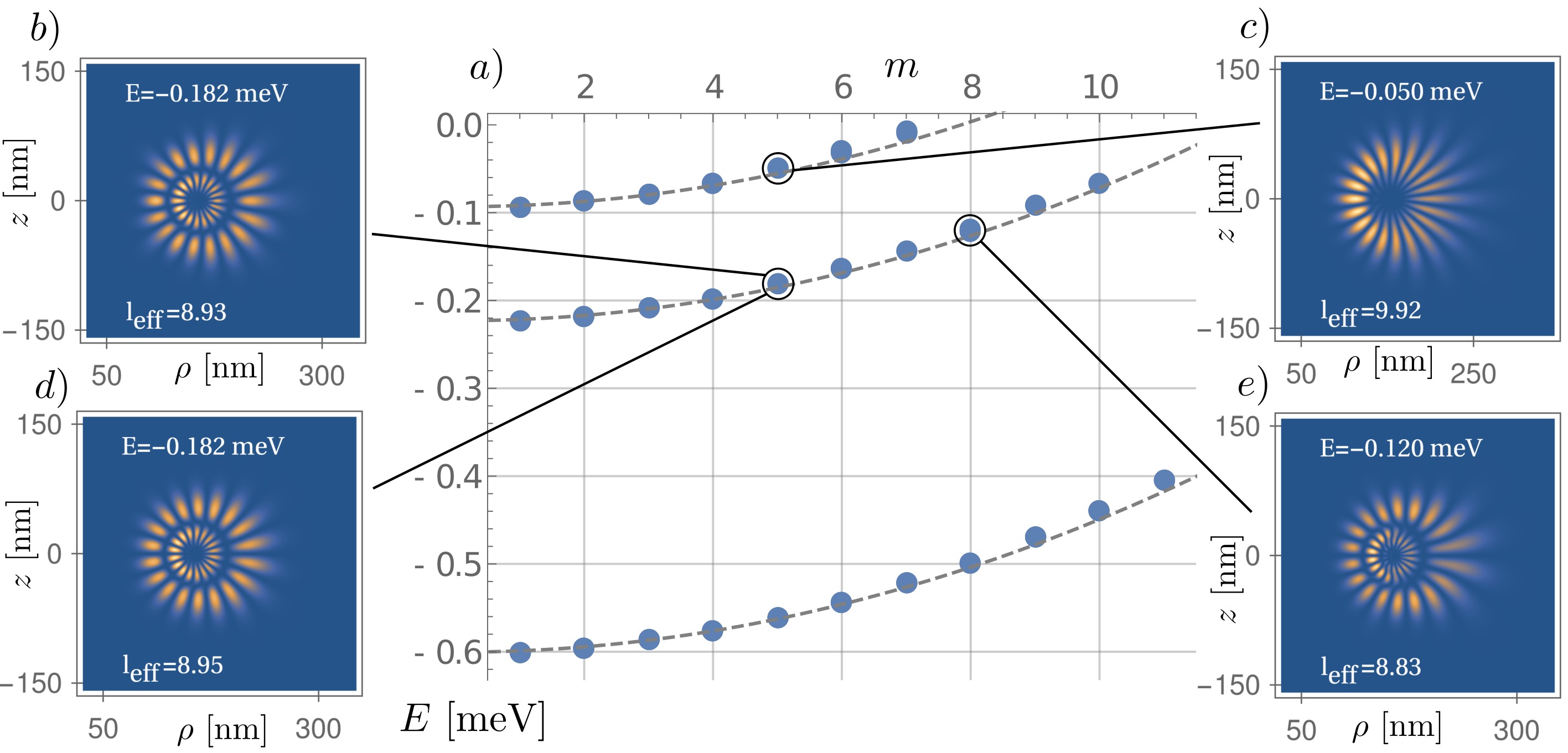}
\caption{Energies $E$ and densities $|u_{m,n}(\rho,z)|^2$ of selected TIS in dependence of the angular momentum number $m$ for the nanoring with $a=3000 $ ($\sim 159 \text{ nm}$) and $b= 30 $ ($\sim 1.59 \text{ nm}$). The blue dots in $a)$ represent the numerically obtained energies (on this scale all energies are two-fold degenerate) which are compared to the approximation based on Eq. (\ref{eqn:energy_m}). The insets $b)$-$e)$ depict the density $|u_{m,n}(\rho,z)|^2$ of the TIS and contain further information on their energy $E$ and their angular momentum $l_\text{eff}$.}
\label{fig:m_spec}
\end{figure}

The above analyzed states correspond hence to a regime where the image potential experienced by the electron is nearly circularly symmetric.    
This situation is comparable to the case of two parallel (infinitely extended) nanotubes sharing an electronic image state which has been analyzed in \cite{segal_tunable_2005}. For sufficiently large tube separations all image states are gerade or ungerade superpositions of single-tube states and can be labeled by a good angular momentum quantum number. The situation changes, however, for smaller tube separations where an increasing asymmetric distortion leads to a collapse of the electronic states onto the tubes.
A similar effect can be observed in the nanoring system for an increasing toroidal angular momentum $m$. This is illustrated in Fig. \ref{fig:m_spec} where we follow the energies and densities of the six energetically highest TIS presented in Fig. \ref{fig:denplots_torus}. The densities can be compared to Fig. \ref{fig:denplots_torus} and illustrate the impact of the additional centrifugal potential. As can be seen an increasing $m$ distorts the circular symmetry of the TIS. For high $m$ this distortion destabilizes the system such that the wave functions are less well-separated from the surface but also less well bound to the nanoring, see Fig. \ref{fig:m_spec} $e)$. To a good approximation the energy shift with respect to $m$ of the detached TIS $u_{m,n}(\rho,z)$ can be described by the parabola
\begin{equation}
E_{m,n}\approx E_{1,n} + \frac{m^2-1/4}{2 a^2}
\label{eqn:energy_m}
\end{equation}
which is shown in Fig. \ref{fig:m_spec} $a)$. 
 This is the expected behavior based on Eq. (\ref{eqn:v_eff}) in the limit that the TIS localizes sufficiently well at a mean distance $a$ from the coordinate center and that the state dependent offset $E_{0,n}$ can be approximately replaced by $E_{1,n}$.     

\subsection{Effective radial potential}
\label{sec:effective_radial_potential}
In a regime where the image potential is almost circularly symmetric around the tube center $\{\rho,z\}=\{a,0\}$, it is beneficial to analyze instead of the full image potential the effective radial potential
\begin{equation}
V^\text{eff}_\text{l,m}(\rho)= \frac{l^2-\frac{1}{4}}{2|\rho-a|^2}+ V^\text{eff}_\text{m}(\rho,0).
\label{eqn:veff_rad}
\end{equation}
$V^\text{eff}_\text{l,m}(\rho)$ includes an additional centrifugal barrier associated to the angular momentum $l$. It can be viewed as an approximation for the potential energy associated to the degree of freedom describing radial motion with respect to the tube center. This approximation becomes accurate for states localized sufficiently close to the tube surface where the potential $V_m^\text{eff}(\rho,z)$ is circularly symmetric. In a more sophisticated treatment small anisotropies could be taken into account by separating the circular and radial motion adiabatically and analyzing instead of (\ref{eqn:veff_rad}) a resulting set of appropriately circularly averaged adiabatic potential curves.

\begin{figure}[h]
\includegraphics[width=0.49 \textwidth]{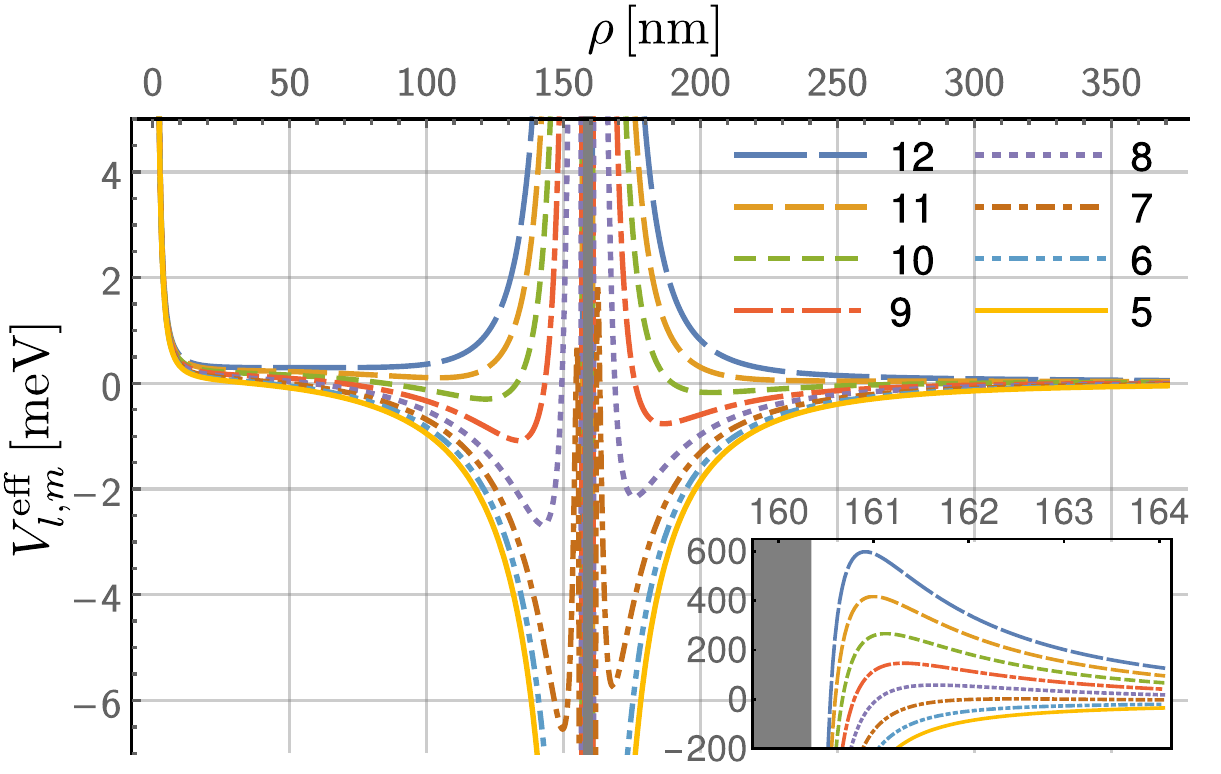}
\caption{The effective radial potential $V^\text{eff}_\text{l,m}(\rho)$ from Eq. (\ref{eqn:veff_rad}) for the nanoring with major radius $a=3000$ ($\sim 159 \text{ nm}$) and minor radius $b=30$ ($\sim 1.59 \text{ nm}$). The angular momenta are $m=1$ and $ 5 \leq l \leq 12$. The inset depicts these curves close to the nanoring surface. In both plots the inner region of the nanoring is marked in gray color.}
\label{fig:veff}
\end{figure}

In Fig. \ref{fig:veff} the radial potential from (\ref{eqn:veff_rad}) is depicted for the exemplary nanoring ($a=3000$, $b=30$) for an angular momentum $m=1$ and effective angular momenta $l$ ranging from 5 to 12.
In the figure all curves with $ 7 \leq l \leq 10$ possess potential wells that are separated by centrifugal barriers from the nanoring surface. As expected from the ring geometry these wells are non-symmetric with respect to reflections at the axis $\rho=a$. Further asymmetry stems from the centrifugal potential associated to $m$ which leads to the divergent behavior for $\rho \to 0$. The properties of the wells can be compared to the exact solutions from Fig. \ref{fig:denplots_torus}. For example, the curve with $l=9$ possesses potential wells at $\rho\approx 130 \text{ nm} $ and  $\rho \approx 190 \text{ nm}$ with depths of approximately 1 meV. These values agree well with the energies and the radial extent of the four TIS possessing $l_\text{eff} \approx 9$ in Fig. \ref{fig:denplots_torus}. Their radial wave functions can be interpreted as the lowest and first exited radial states in these wells. Similar agreement can be found for the other states in Fig. \ref{fig:denplots_torus}. Furthermore, the radial potential predicts that TIS with $l_\text{eff}> 10 $ and $l_\text{eff} < 7$ are either not bound to the ring or are not detached from its surface which agrees with our numerical results. 

In a regime where the circular and the radial motion are approximately separable, the effective potential $V^\text{eff}_\text{l,m}(\rho)$ is consequently a useful tool to identify angular momenta $l_\text{eff}$ leading to detached TIS and to specify their properties. In the following we will employ $V^\text{eff}_\text{l,m}(\rho)$ to estimate general criteria for the existence of detached TIS at nanorings in a regime where the impact of $m$ is small.
A general framework to study the conditions for which infinite metal cylinders support detached TIS has been provided in \cite{segui_tubular_2012}. In particular it has been shown that an infinite cylinder of radius $b$ possesses only detached effective potential wells if the dimensionless parameter $\lambda_l=(l^2-1/4)a_0/2b$ lies close to $0.9$, where $a_0$ denotes the Bohr radius. This is a useful criterion to estimate properties for cylinders having different geometries. 

By performing a similar scaling analysis for the potential (\ref{eqn:v_eff}) one can derive analogous criteria for the nanoring. For the exemplary nanoring one finds e.g. that potential wells exists for $\lambda_l \approx 1$ and one can conclude that this is also true for all other nanorings having the same ratio $b/a=0.01$. We computed critical $\lambda_l$ also for nanorings with other ratios $ 0.001 \leq b/a \leq 0.04$ and found $\lambda_l$ lying always between 0.7 and 1.1.

For larger $b/a$, however, the potential wells become so broad, that their identification becomes increasingly difficult. This behavior can be understood by means of the local power-law scaling of the nanoring image potential that depends on the ratio $b/a$, see Sec. \ref{sec:image_pot_analysis}.       
A basic calculation shows that an effective potential of the form $V^\text{eff}(\rho)= c_1 \rho^{\alpha} + c_2\rho^{-2}$, where $c_1$, $c_2$ and $\alpha$ are constants, possesses minimal extremal values only for $\alpha<-2$. This is for example the reason why there exist no high angular momentum image states around spherical graphite particles \cite{gumbs_strongly_2014}. This criterion can also be applied locally and we conclude from Fig. \ref{fig:exponent} for the nanoring that potential wells supporting detached TIS exist only if $b/a<0.1$, i.e. for sufficiently thin tori.

\section{Conclusions}
\label{sec:conclusions}
In the present work we constructed an analytical expression for the electrostatic image potential between a point charge and a neutral metallic ring. We analyzed this potential and demonstrated that it supports image states ``circulating'' the torus which can be viewed as a curvilinear generalization of tubular image states around nanocylinders \cite{granger_highly_2002}. In contrast to the latter the angular momentum $l$ shielding the states from the torus surface is not conserved and the states are not circularly symmetric. The second angular momentum $m$ (associated to the motion along the nanoring) is conserved and can be used to tune the energy and also to some extent the shape of the image states. Too high $m$ will, however, couple the different $l$ states too strongly and destabilize the system.
The main features of TIS can be explained by means of an effective radial potential including a centrifugal barrier associated to $l$. By analyzing the scaling properties of this potential we argued that centrifugally detached high angular momentum image states exist only around sufficiently thin nanorings having aspect ratios smaller than $b/a=0.1$. However, it has to be expected that nanorings (like nanotubes) also support low angular momentum TIS states probing the surface, whose investigation might be an interesting subject for future studies.

In the last decades there has been a tremendous progress in the development of nanofabrication techniques \cite{gates_new_2005} and the production of almost perfectly ring-shaped silver nanowires with dimensions $a\approx 5 \mu m$ and $b\approx 50 nm$ is already possible \cite{zhou_crystal_2009}. Compared to experimentally realized TIS around straight cylindrical structures the additional curvature of nanorings could offer interesting new applications. For example the repulsive mutual Coulomb interaction of two or more simultaneously excited electrons at a nanoring could lead to special equilibrium configurations where each electron resides at a particular angle along the ring.
The here presented results are a first exploration of TIS at curved nanowires and should be of relevance for all future studies and applications of image states at even more complex-shaped nanostructures. 

%%%%%%%%%%%%%%%%%%%%%%%%%%%%%%%%%%%%%%%%%%%%%%%%%%%%%%%%%%%%%%%%%%%%%%%%%%%%%
\begin{acknowledgments}
C.F. gratefully acknowledges a scholarship by the Studienstiftung des deutschen Volkes.
\end{acknowledgments}
%%%%%%%%%%%%%%%%%%%%%%%%%%%%%%%%%%%%%%%%%%%%%%%%%%%%%%%%%%%%%%%%%%%%%%%%%%%%%
%\bibliographystyle{unsrt}
%\bibliographystyle{apsrev4-1}
%\bibliography{nano}
%\begin{thebibliography}{}

 % \end{thebibliography}
%%%%%%%%%%%%%%%%%%%%%%%%%%%%%%%%%%%%%%%%%%%%%%%%%%%%%%%%%%%%%%%%%%%%%%%%%%%%%
\appendix
\section{Derivation of the image potential for the torus}
\label{sec:imagepotential_derivation}
Our aim is to construct the electrostatic potential $\Phi_\text{tot}(\vec{r};\vec{r}_e)$ outside a neutral metallic torus in the presence of an electron at position $\vec{r}_e$. This setup is very similar to the problem of a point charge in the proximity of an isolated, neutral and conducting sphere \cite{jackson_classical_1999} and can be solved by a related approach. To this aim we construct $\Phi_\text{tot}$ as a linear superposition of the form
\begin{equation}
\Phi_\text{tot}(\vec{r};\vec{r}_e)=\Phi^\text{g}_\text{tot}(\vec{r};\vec{r}_e)+ \alpha(\vec{r_e}) \Phi^1(\vec{r}) 
\label{eqn:pot_torus_ansatz}
 \end{equation}
where $\Phi^\text{g}_\text{tot}$ is the potential around a grounded metallic torus in the presence of the electron and $\Phi^1$ is the particular potential of a charged metallic torus with the constant surface potential $\Phi^1|_A=1$. The factor $\alpha(\vec{r}_e)$ needs to be determined from the condition that the total charge in the inner torus region vanishes, i.e. that the induced potential is source free. 

The solution $\Phi^\text{g}_\text{tot}$ has been presented in \cite{scharstein_electrostatic_2005}. It can be written as $\Phi^\text{g}_\text{tot}(\vec{r};\vec{r}_e)=\Phi_\text{ind}^\text{g}(\vec{r};\vec{r}_e)-1/|\vec{r}-\vec{r}_e|$ where the induced potential reads in toroidal coordinates  
\begin{align}
\Phi^\text{g}_\text{ind}(\vec{r};\vec{r}_e)&=\frac{1}{\pi c} (\cosh \xi - \cos \eta)^{1/2} (\cosh \xi_e - \cos \eta_e)^{1/2} \nonumber \\
&\cdot \sum_{n=0}^\infty \sum_{m=0}^\infty \epsilon_n \epsilon_m (-1)^m \frac{\Gamma(n-m+1/2)}{\Gamma(n+m+1/2)} \frac{\cos [m (\varphi -\varphi_e)] \cos \left[n(\eta-\eta_e)\right]}{P_{n-1/2}^m(\cosh \xi_0)} \nonumber \\
&\cdot Q_{n-1/2}^m(\cosh \xi_0) P_{n-1/2}^m(\cosh \xi_e) P_{n-1/2}^m(\cosh \xi) .
\end{align}
Here $\xi_e$, $\eta_e$, $\varphi_e$ are the toroidal coordinates of the electron.
For large $r \to \infty$, i.e. $\{\xi,\eta\}\to\{0,0\}$, one finds
\begin{equation}
\Phi_\text{ind}^\text{g}\to\frac{\sqrt{2}}{\pi r}(\cosh \xi_e - \cos \eta_e)^{1/2} \sum_{n=0}^\infty \epsilon_n \frac{Q_{n-1/2}(\cosh \xi_0)}{P_{n-1/2}(\cosh \xi_0)}  P_{n-1/2}^m(\cosh \xi_e) \cos(n\eta_e) .
\end{equation}
By comparing this expression to the multipole expansion $\Phi_\text{ind}(\vec{r})= q_\text{ind}/r + \mathcal{O}(1/r^2)  $ and by substituting (\ref{eqn:sum1}) one can identify the induced positive monopole
\begin{equation}
q_\text{ind}(\vec{r}_e)=\frac{\sqrt{2}}{\pi}(\cosh \xi_e - \cos \eta_e)^{1/2} s_1(\xi_e,\eta_e) .
\end{equation}

This monopole needs now to be compensated by the potential $\alpha(\vec{r}_e) \Phi^1(\vec{r})$.
The therefore required solution $\Phi^1(\vec{r})$ is stated in \cite[p.239]{smythe_static_1989} \cite[p.73]{smythe_solutions_1974}. By employing Whipple's transformation for Legendre functions and (\ref{eqn:sum1}) it can be expressed in toroidal coordinates as 
\begin{align}
\Phi^1(\vec{r}) &=\frac{\sqrt{2}}{\pi} \sqrt{\cosh \xi -\cos \eta} \sum_{n=0}^\infty \epsilon_n \frac{Q_{n-1/2}(\cosh \xi_0)}{P_{n-1/2}(\cosh \xi_0)} P_{n-1/2}(\cosh \xi) \cos (n \eta) \nonumber \\
&=\frac{\sqrt{2}}{\pi} \sqrt{\cosh \xi -\cos \eta} \, s_1(\xi,\eta) .
\label{eqn:potential_insulated_charged_torus}
\end{align}
From the long-range behavior of (\ref{eqn:potential_insulated_charged_torus}) one can derive  
the capacitance of the torus $C= 2c/\pi \cdot s_1(0,0)$ which depends only on the geometry of the torus and links the induced charge to the surface potential \cite{smythe_static_1989,smythe_solutions_1974}. Due to the linearity of the Laplace equation the solution $\alpha(\vec{r}_e) \Phi^1(\vec{r})$ corresponds to a torus with surface potential $\alpha(\vec{r}_e)$ carrying a charge $C \alpha(\vec{r}_e)$.
Consequently, by setting  $\alpha(\vec{r}_e)=- q_\text{ind}(\vec{r}_e)/C$ one obtains with (\ref{eqn:pot_torus_ansatz}) and (\ref{eqn:potential_insulated_charged_torus}) the total potential 
\begin{align}
\Phi_\text{tot}(\vec{r};\vec{r}_e)&= \Phi^\text{g}_\text{tot}(\vec{r};\vec{r}_e) -\frac{q_\text{ind}(\vec{r}_e)}{C}\Phi^1(\vec{r};V) \nonumber \\
&= \Phi^\text{g}_\text{tot}(\vec{r};\vec{r}_e) - \frac{1}{\pi c}(\cosh \xi_e -\cos \eta_e)^{1/2} (\cosh \xi -\cos \eta)^{1/2} \frac{s_1(\xi_e,\eta_e) s_1(\xi, \eta)}{s_1(0,0)}  \ .
\label{eqn:torus_phi_total}
\end{align}
The image potential is consequently given by
\begin{align}
V_\text{im}(\vec{r})&= -\frac{1}{2} \left(\Phi_\text{ind}^\text{g}(\vec{r};\vec{r}) -\frac{q_\text{ind}(\vec{r}_e)}{C} \Phi^1(\vec{r})\right) \nonumber \\
&= \frac{1}{2 \pi c} (\cosh \xi -\cos \eta) \left[ \frac{s_1^2(\xi,\eta)}{s_1(0,0)}-s_3(\xi,\eta) \right]  \ .
\end{align}

\end{document}